\documentclass[12pt, epsf]{article}
\usepackage[cp1251]{inputenc}
\usepackage[russian]{babel}
\begin{document}
\large
\par
{\bf Majorana neutrino. Is double neutrinoless beta decay possible
in the framework of the weak interactions? How to prove that
neutrino is Majorana particle.}
\begin{center}
\par
\vspace{0.3cm} Beshtoev Kh. M. (beshtoev@cv.jinr.ru)
\par
\vspace{0.3cm} Joint Institute for Nuclear Research, Joliot Curie
6, 141980 Dubna, Moscow region, Russia.
\end{center}
\vspace{0.3cm}

\par
Abstract

\par
Usually it is supposed that Majorana neutrino produced in the
superposition state $\chi_L = \nu_L + (\nu_L)^c$ and then follows
the neutrinoless double beta decay. But since weak interactions
are chiral invariant then the neutrino at production has definite
helicity (i.e.,  $\nu_L$ and $(\nu_L)^c$ neutrinos are separately
produced and then neutrino is not in the superposition state).
This helicity cannot change after production without any external
interactions. Thus we see that for unsuitable helicity the
neutrinoless double $\beta$ decay is not possible even if neutrino
is a Majorana particle. Also transition of Majorana neutrino into
antineutrino at their oscillations is forbidden since helicity in
vacuum holds. Then only possibility to prove that neutrino is a
Dirac but not Majorana particle is detection transition of $\nu_L$
neutrino into (sterile) antineutrino $\bar\nu_R$ (i.e., $\nu_L \to
\bar\nu_R$) at neutrino oscillations. Transition Majora neutrino
$\nu_L$ into $(\nu_R)^c$ (i.e., $\nu_L \to (\nu_R)^c$) at
oscillations is unobserved since it is supposed that mass of
$(\nu_R)^c$ is very big.\\
\par
\noindent
PACS numbers: 14.60.Pq; 14.60.Lm

\section{Introduction}

\par
\hspace{0.4cm} The equation for particle with spin $\frac{1}{2}$
was first time formulated by Dirac \cite{1 dirac} in 1928.
Afterwards it turned out, that this representation was adequate to
describe neutral and charged fermions, i.e., fermions are Dirac
particles. Later Majorana found an equation \cite{2 maj} for
fermion with spin $\frac{1}{2}$. Then it became clear that this
fermion could be only a neutral particle since in one
representation there is a joint of the particle and
antipartic\-le.
\par
In 50-es years of the 20 century the Majorana neutrino study was
very wide \cite{3 pauli}. Letter it stopped.
\par
The suggestion that in analogy with $K^{o},\bar K^{o}$
oscillations there could be neutrino-antineutrino oscillations (
$\nu \rightarrow \bar \nu$), was considered by Ponte\-corvo
\cite{4 pontecorvo} in 1957. It was subsequently considered by
Maki et al. \cite{5 maki} and Pontecorvo \cite{6 pontecorvo2} that
there could be mixings (and oscillations) of neutrinos of
different flavors (i.e., $\nu _{e} \rightarrow \nu _{\mu }$
transitions). A complete consideration of Majorana neutrino
oscillations was given in \cite{7 pontecorvo}.
\par
A posteriori Majorana neutrino can be introduced in two ways:
\par
1. To suppose that Majorana neutrino is superpositions of $\nu_L$
and $(\nu_L)^c$ \cite{2 maj}, \cite{8 bil}
$$
\chi = \nu_L + (\nu_L)^c \quad (\nu_L)^c \equiv C \bar \nu_L^T .
$$
Here arises a question: is neutrino production possible by weak
interactions in this superposition state? Unfortunately, this wave
function is normalized on two. The definition of Majorana neutrino
normalized on one was given in \cite{9 boehm} and then
$$
\chi = \frac{1}{\sqrt{2}} (\nu_L + (\nu_L)^c) .
$$
Experimental consequences of this definition of Majorana neutrino
was considered in \cite{10 besh}.
\par
2. To suppose that the standard definition of Majorana neutrino
$$
\chi = \nu_L + (\nu_L)^c ,
$$
is a formal recording and it has no physical realization. Such
supposition is a direct consequence of the weak interactions where
neutrino can be produced only with definite helicity), it means
that neutrino cannot be produced in the superposition state. The
weak interactions cannot produce neutrino in the mixing helicity
states since these interac\-tions are chiral invariant (i.e. the
weak interactions cannot produce neutrino in the superposition
state). From all known experiments the neutri\-nos in weak
interactions are produced with definite spirality (helicity)
\cite{11 data}. It is necessary to remark that this result is a
right consequence of weak interactions which cannot produce
neutrino $\chi$ which is a superposition of $\nu_L$ and
$(\nu_L)^c$ neutrinos.

\par
Review of tasks related the problem of distinguishing Dirac and
Majorana neutrinos was considered in \cite{12 zralek}, with a big
quantity of references.
\par
At the present time after detection that there are transitions
between different types of neutrino a very important problem
appears: is neutrino a Dirac or Majorana particle?
\par
This work is purposed to clarify the questions connected with
Majorana neutrino and therefore it should be discussed.

\par
\section{Majorana neutrino}

\par
Gamma matrices have the following form (we follow notation in
\cite{8 bil}):
$$
\gamma^o = \left(\begin{array}{cc} 0 & -1 \\
-1 & 0 \end{array}\right) ,  \gamma^i =
\left(\begin{array}{cc} 0 & \sigma_i \\
-\sigma_i & 0 \end{array}\right) ,  \gamma^5 =
-i \gamma^o \gamma^1 \gamma^2 \gamma^3 =
\left(\begin{array}{cc} -1 & 0 \\
0 & 1 \end{array}\right) , \eqno(1)
$$
where $i = 1 \div 3$ and $\sigma_i$ are Pauli matrices. And
$$
\gamma^\mu \gamma^\nu + \gamma^\nu \gamma^\mu = 2 g^{\nu \mu} ,
\quad \gamma^\mu \gamma^5 + \gamma^5 \gamma^\mu = 0 , \eqno(2)
$$
where $\mu, \nu \div 0,1,2,3$, $g^{\nu \mu} = 0$ if $\nu \neq \mu$
and $g^{\nu \mu} = (1, -1, -1, -1)$ if $\nu = \mu$.

\par
Usually a Majorana neutrino (antineutrino) is connected with Dirac
antineutrino $\bar \nu_L, \bar \nu_R$ in the following manner (it
is necessary to draw attention that in this section of this work
the notation of work \cite{8 bil} it is used):
$$
(\nu_L)^c \equiv C \bar \nu_L^T, \quad (\nu_R)^c \equiv C \bar
\nu_R^T , \eqno(3)
$$
where $C$ is a charge-conjugation matrix, and this matrix
satisfies the conditions ($C \sim \gamma^4 \gamma^2$)
$$
C \gamma^{\mu T} C^{-1} = - \gamma^\mu, \quad C^{+} C = 1, \quad
C^{T} = -C . \eqno(4)
$$
Using (3), (4) we can obtain that
$$
\overline{(\nu_L)^c} = - \nu_L^T C^{-1}, \quad
\overline{(\nu_R)^c} = - \nu_R^T C^{-1} .  \eqno(5)
$$
Now it is necessary to find out what type of fermions are the
above Majorana $ (\nu_L)^c, (\nu_R)^c$ neutrinos. For this reason
to these states projection operators are applied
$$
\frac{1}{2}(1 - \gamma^5) (\nu_L)^c = C [\bar \nu_L \frac{1}{2} (1
- \gamma^5)]^T , \eqno(6)
$$
where we used
$$
C^{-1} \gamma^5 C = \gamma^{5 T}. \eqno(7)
$$
Since $\bar \nu_L \frac{1}{2} (1 - \gamma^5) = \bar \nu_L$, then
from (6) we get
$$
\frac{1}{2}(1 - \gamma^5) (\nu_L)^c = (\nu_L)^c . \eqno(8)
$$
So we come to a conclusion that $(\nu_L)^c$ neutrino is a
right-sided neutrino. In the similar way we can get
$$
\frac{1}{2}(1 + \gamma^5) (\nu_R)^c = (\nu_R)^c , \eqno(9)
$$
that neutrino $(\nu_R)^c$ is a left-sided neutrino. So, instead of
four neutrino (fermion) states in the case of Dirac fermions $\bar
\nu_R, \bar \nu_L, \nu_R, \nu_L$ in the Majorana case four
neutrino states $(\nu_R)^c, \nu_R, (\nu_L)^c, \nu_L$ appear here.
\par
Majorana equation for neutrino is \cite{2 maj}
$$
i (\widehat{\sigma}^\mu d_\mu) \nu_R - m^M_R \epsilon \nu^{*}_R =
0,
$$
$$
i (\widehat{\sigma}^\mu d_\mu) \nu_L - m^M_L \epsilon \nu^{*}_L =
0, \eqno(10)
$$
where $\widehat{\sigma}^\mu \equiv (\sigma^0,
\overrightarrow{\sigma})$, $\sigma^\mu \equiv (\sigma^0, -
\overrightarrow{\sigma})$, \overrightarrow{\sigma} is Pauli
matrices,
$$
\epsilon = \left ( \begin{array}{cc} 0, 1 \\
-1, 0 \end{array} \right ).
$$
These equations describe two completely different neutrinos with
masses $m^M_R$ and $m^M_L$ which do not posses any additive
numbers and neutrinos are their own antineutrinos (i.e., particles
differ from antiparticles only in spin projections). Now it is
possible to introduce the following two Majorana neutrino states:
$$
\begin{array}{c} \chi_L = \nu_L + (\nu_L)^c \\
\chi_R = \nu_R + (\nu_R)^c \end{array} . \eqno(11)
$$
Formally the above Majorana equation (10) can be rewritten in the
form
$$
(\gamma^\mu \partial_\mu + m) \chi (x) =0 , \eqno(12)
$$
with the Majorana condition ($\chi \equiv \chi_{L R}$)
$$
C \bar \chi^T (x) = \xi \chi (x) , \eqno(13)
$$
where $\xi$ is a phase factor ($\xi = \pm 1$) and $C$ is matrix
(2).
\par
It is necessary to stress that
$$
\bar \chi (x) \gamma^\mu \chi (x) = 0, \eqno(14)
$$
i.e., vector currant of Majorana neutrino is equal to zero.
\par
At the beginning in the Dirac representation we have two states -
neutrino state $\Psi_L$ and antineutrino state $\bar \Psi_L$, then
by using Majorana condition (3) we come to two new neutrino
states: $\Psi_L$ and $(\Psi_L)^c$. The question is: can we
construct one Majorana neutrino state from these two states, as it
was done in the above consideration while obtaining exp. (11),
(12). As a matter of fact it is necessary to introduce two
Majorana neutrino states:
$$
\begin{array}{c} \chi_{1 L} = \frac{1}{\sqrt{2}} (\nu_L + (\nu_L)^c) \\
\chi_{2 L} = \frac{1}{\sqrt{2}} (- \nu_L + (\nu_L)^c) \end{array}
, \eqno(15)
$$
as it was considered in \cite{9 boehm} (i.e., these must be two
Majorana neutrino states but not one Majorana neutrino state).
\par
As it is noted above the neutrino states $\chi_{L R} (x)$ in (11)
are normalized on 2. It is well known that state functions of
particles must be normalized on one. It is well seen in the
example of neutrino antineutrino mixing considered by B.
Pontecorvo \cite{4 pontecorvo}, where two normalized states $\nu_1
= \frac{1}{\sqrt{2}} (\bar \nu_e + \nu_e)$, $\nu_2 =
\frac{1}{\sqrt{2}} (\bar \nu_e - \nu_e)$, appear. It is clear that
when formulating Majorana neutrino the second state in (15) was
not taken into account.
\par
Majorana mass lagrangian can be written in the following form:
$$
{\cal L}^M = -\frac{1}{2} \overline{(\nu_L)^c} m^M_L \nu_L
-\frac{1}{2} \overline{(\nu_R)^c} m^M_R \nu_R  + H. c.  \eqno(16)
$$

\par
Lagrangian ${\cal L}^M_I (\chi ...)$ interaction of Majorana
electron neutrinos with electrons usually has the following form:
$$
{\cal L}^M_I (e, \chi_L) = \frac{i g}{2 \sqrt{2}} \bar e_L (x)
\gamma^\mu \chi_L (x) W^{-}_\mu + H. c. \eqno(17)
$$
It is necessary to remark that lagrangian (17) for the Majorana
neutrino interaction, in contrast to the lagrangian for the Dirac
neutrino interac\-tion in the electroweak model \cite{13 salam},
is not invariant relatively to the weak isospin transformation
(Majorana neutrino has no zero weak isospin). Also a relation
analogous to the Gell-Mann-Nishigima in this case is absent.
Besides here the global gauge invariance is absent since the
Majorana neutrino state is a superposition of neutrino and
antineutrino. Most sorrowful is that in this lagrangian the local
gauge invariance is violated (the local gauge invariance can be
fulfilled in the case when there is a Dirac particle and
antiparticle). Hardly violation of local gauge invariance by hand
has a since. This violation requires a serious substantiation.
Neutrinos are right produced together with leptons and quarks at
the energies where electroweak model works very well. So, a
supposition that neutrinos are Majorana neutrinos is absolutely
unfounded. For Majorana neutrinos there is only one possibility:
if in reality in the nature the Majorana neutrinos exist then at
very hight energies, if local gauge invariance is violated, then
Dirac neutrinos can be converted into Majorana neutrinos.

\par
\section{Is neutrinoless double $\beta$ decay
possible if neutrinos are Majorana neutrinos?}

\par
A reaction with a double beta decay with two electrons
$$
(Z, A) \rightarrow (Z+2, A) + e^{-}_1 + e^{-}_2 + \bar \nu_{e 1} +
\bar \nu_{e 2} ,
$$
is possible if $M_A (Z,A) > M_A (Z+2,A)$.
\par
In analogy with the electron double beta decay there can be
reactions with double neutrino radiation by electron capture or
positron radiation
$$
(Z, A) + e^{-}_1 + e^{-}_2 \rightarrow (Z-2, A) + \nu_{e 1}
+\nu_{e 2} ,
$$
if $M_A (Z,A) > M_A (Z-2,A) + 2 \Delta$;

$$
(Z, A) + e^{-}_1 \rightarrow (Z-2, A) + e^{+}_{2} + \nu_{e 1} +
\nu_{e 2} ,
$$
if $M_A (Z,A) > M_A (Z-2,A) + 2 m_e + \Delta$;

$$
(Z, A) \rightarrow (Z-2, A) + e^{+}_1 + e^{+}_2 + \nu_{e 1} +
\nu_{e 2} ,
$$
if $M_A (Z,A) > M_A (Z-2,A) + 4 m_e$, where $\Delta$ is the
binding energy of electron.
\par
If neutrino is a Majorana particle ($\chi_L = \nu_L + (\nu_L)^c$)
then the following neutrinoless double beta decays are possible:
$$
(Z, A) \rightarrow (Z+2, A) + e^{-}_1 + e^{-}_2 ,
$$
if $M_A (Z,A) > M_A (Z+2,A)$;
\par
\noindent
and
$$
(Z, A) + e^{-}_1 + e^{-}_2 \rightarrow (Z-2, A) ,
$$
if $M_A (Z,A) > M_A (Z-2,A) + 2 \Delta$;

$$
(Z, A) + e^{-}_1 \rightarrow (Z-2, A) + e^{+}_{2} ,
$$
is possible if $M_A (Z,A) > M_A (Z-2,A) + 2 m_e + \Delta$;

$$
(Z, A) \rightarrow (Z-2, A) + e^{+}_1 + e^{+}_2 ,
$$
if $M_A (Z,A) > M_A (Z-2,A) + 4 m_e$;
\par
The lepton part of the amplitude of the two neutrino decay has the
following form \cite{9 boehm}, \cite{14 vergados}:
$$
\bar e(x) \gamma_\rho \frac{1}{2}(1 \pm \gamma_5) \nu_j  \bar e(y)
\gamma_{\sigma} \frac{1}{2} (1 \pm \gamma_5)  \nu_k (y) .
\eqno(18)
$$
After substituting of the neutrino propagator and its integrating
on the momentum of virtual neutrino, the lepton amplitude gets the
following form:
$$
-i \delta_{jk} \int \frac{d^4 q}{(2 \pi)^4} \frac{e^{- i q
(x-y)}}{q^2 - m_j^2} \bar e(x) \gamma_\rho \frac{1}{2}(1 \pm
\gamma_5) (q^\mu \gamma_\mu + m_j) \frac{1}{2} (1 \pm \gamma_5)
\gamma_{\sigma} e(y). \eqno(19)
$$
If we use the following expressions
$$
\ \frac{1}{2}(1 - \gamma_5) (q^\mu \gamma_\mu + m_j) \frac{1}{2}
(1 - \gamma_5)  = m_j \frac{1}{2} (1 - \gamma_5),  \eqno(20)
$$
$$
 \frac{1}{2}(1 - \gamma_5) (q^\mu \gamma_\mu + m_j)
\frac{1}{2} (1 + \gamma_5)  = q^\mu \gamma_\mu \frac{1}{2} (1 +
\gamma_5),  \eqno(21)
$$
then we see that in the case when there are only left currents
(expression (20)) we get a deposit only from the neutrino mass
part,
$$
-i \delta_{jk} \int \frac{d^4 q}{(2 \pi)^4} \frac{e^{- i q
(x-y)}}{q^2 - m_j^2} \bar e(x) \gamma_\rho \frac{1}{2}(1 -
\gamma_5) (q^\mu \gamma_\mu + m_j) \frac{1}{2} (1 - \gamma_5)
\gamma_{\sigma} e(y) =
$$
$$
= -i \delta_{jk} \int \frac{d^4 q}{(2 \pi)^4} \frac{e^{- i q
(x-y)}}{q^2 - m_j^2} \bar e(x) \gamma_\rho m_j \frac{1}{2} (1 -
\gamma_5) \gamma_{\sigma} e(y), \eqno(22)
$$
while at presence of the right currents (expression (21))
$$
-i \delta_{jk} \int \frac{d^4 q}{(2 \pi)^4} \frac{e^{- i q
(x-y)}}{q^2 - m_j^2} \bar e(x) \gamma_\rho \frac{1}{2}(1 -
\gamma_5) (q^\mu \gamma_\mu + m_j) \frac{1}{2} (1 + \gamma_5)
\gamma_{\sigma} e(y) =
$$
$$
= -i \delta_{jk} \int \frac{d^4 q}{(2 \pi)^4} \frac{e^{- i q
(x-y)}}{q^2 - m_j^2} \bar e(x) \gamma_\rho q^\mu \gamma_\mu
\frac{1}{2} (1 + \gamma_5) \gamma_{\sigma} e(y) , \eqno(23)
$$
the amplitude includes in the neutrino propagator  term
proportio\-nal to four momentum $q$.
\par
We see that if Majorana neutrino is a superposition of $\nu_L$ and
$(\nu_L)^c$ then the neutrinoless double beta decay will be take
place.
\par
Now we come to another consideration: is neutrinoless double
$\beta$ decay possible if neutrinos are Majorana neutrinos in the
case when it is taken into account that in the weak interactions
neutrino is produced in definite spirality (helicity)?
\par
Our consideration begins with an example of $\pi^\pm$ decays:
$$
\begin{array}{c} \pi^{+} \to e^{+} + \nu_e \\
\pi^{-} \to e^{-} + \bar\nu_e
\end{array} . \eqno(24)
$$
If neutrino is a Dirac particle then $\nu_e$ is described by wave
function $\Psi_{eL}$ and $\bar \nu_e$ is described by wave
function $\bar \Psi_{eL}$. If neutrino is a Majorana particle then
$\nu_e$ is described by wave function $\Psi_{eL}$ and $\bar \nu_e$
is described by wave function $(\Psi_{eL})^c$. If Majorana
neutrino is $\chi_e = \Psi_{eL} + (\Psi_{eL})^c$ as usually it is
supposed, then in the above two processes the Majorana electron
neutrino state $\Psi_{eL}$ are produced in the first case and
Majorana electron neutrino $(\Psi_{eL})^c$ in the second case are
produced.
\par
So, since in the weak interactions neutrino can be produced only
in definite spirality (helicity) state but not in with spirality
(helicity) a mixing state as it is supposed in exp. (11), then
neutrino will be produced in state $\nu_L$ or $(\nu_L)_c$
\par
For example, the double nuclear beta decay with electron radiation
takes place in the following double transition:
$$
(Z, A) \to (Z+1, A) + e^{-}_1 + \bar \nu_e ,
$$
$$
(Z+1, A) \to (Z+2, A) + e^{-}_2 + \bar \nu_e . \eqno(26)
$$
If neutrino is a Majorana neutrino then we can rewrite the above
two expressions in the following form:
$$
(Z, A) \to (Z+1, A) + e^{-}_1 + \bar \nu_e ((\Psi_{eL})^c) \to
$$
$$
e^{-}_1 + \bar \nu_e ((\Psi_{eL})^c) + (Z+1, A) \to (Z+2, A) +
e^{-}_1 + e^{-}_2 , \eqno(27)
$$
but for realization of the second process it is necessary to have
neutrino state $\Psi_{eL}$
$$
\nu_e (\Psi_{eL}) + (Z+1, A) \to (Z+2, A) + e^{-}_2 , \eqno(28)
$$
but not $(\Psi_{eL})^c$ neutrino state (it is clear that if weak
interactions radiate neutrino in the superposition state then this
process can be realized). Since the first reaction produces
$(\Psi_{eL})^c$ but not the $\Psi_{eL}$ neutrino state then their
convolutions
$$
\widehat{(\Psi_{eL})^c (\Psi_{eL})^c} = 0 , \quad
\widehat{\Psi_{eL} \Psi_{eL}} = 0 ,
$$
are zero and therefore the above second reaction is forbidden:
$$
\nu_e (\Psi_{eL}) + (Z+1, A) \not\rightarrow (Z+2, A) + e^{-}_2 .
\eqno(29)
$$
\par
As mentioned above, usually it is supposed that Majorana neutrino
is produced in the first reaction and then it is absorbed in the
second reaction and then the neutrinoless double beta decay
arises. But in this considered case two Majorana neutrinos are
produced in state $(\Psi_{eL})^c$ (weak interactions are chiral
symmetric interactions and neutrino at production has definite
spirality (helicity) and after production this spirality
(helicity) cannot change without an external interaction), then
here the capture of neutrino produced in the first reaction cannot
take place by the second reaction since to realize this
possibility in the first reaction $\Psi_{eL}$ (Majorana) neutrino
must be emitted and then it can be absorbed in the second reaction
(or reverse process can be realized). So we see that for
unsuitable spirality (helicity) the neutrinoless double $\beta$
decay is not possible even if neutrino is a Majorana particle.

\section{How to prove that neutrino is a Majorana particle?}

So, if neutrino is a Dirac particle then $\nu_L$ and $\bar \nu_L$
neutrino states are produced but if neutrino is a Majorana
particle, the $\nu_L$ and $(\nu_L)^c$ neutrino states are produced
(since weak interactions are chiral invariant then neutrino at
production has definite (helicity) spirality)  therefore the
Majorana neutrino eigenstate $\chi = \nu_L + (\nu_L)^c$ cannot be
realized, i.e., $\nu_L$ and $(\nu_L)^c$ neutrinos are separately
produced. As it was shown above in this case, the neutrinoless
double beta decay is forbidden. Then the question arises: how to
prove that neutrino is a Majorana particle.
\par
If neutrino is a Majorana particle there is no conserved lepton
number then mixing and oscillations between neutrino states will
arise,
$$
\begin{array}{c}
\nu_e = cos \theta  \nu_{1} + sin \theta \nu_{2}           \\
(\nu_e)^c = - sin \theta  \nu_{1} + cos \theta  \nu_{2} ,
\end{array}
\eqno(30)
$$
i.e., the $\nu_L$ and $(\nu_L)^c$ neutrino states are transformed
into superpositions of the $\nu_1$ and $\nu_2$ neutrino states.
Then oscillations between $\nu_L$ and $(\nu_L)^c$ neutrino will
take place and a probability of $P(\nu_e \rightarrow \nu_e)$
transitions is
$$
\begin{array}{c}
P(\nu_e \rightarrow \nu_e, t)  = sin^{2} 2\theta sin^2(t
\pi(m^{2}_{2} - m^{2}_{1}) / 2p)  ,
\end{array}
\eqno(31)
$$
where it is supposed that $p \gg  m_{1}, m_{2}; E_{k} \simeq p +
m^{2}_{k} / 2p$ and $m_{1}, m_{2}$ are masses of $\nu_1, \nu_2$
neutrinos.
\par
In this case at neutrino oscillations the neutrinos can be
transformed into antineutrinos and vice versa. Since neutrino
$\nu_L$ and $(\nu_L)^c$ are produced separately and they have
opposite spiralities (heliscities) then such transitions at
neutrino oscillations in vacuum must be absent. Then only
possibility to prove that neutrino is a Dirac but not Majorana
particle is detection transition of $\nu_L$ neutrino into
(sterile) antineutrino $\bar\nu_R$ (i.e., $\nu_L \to \bar\nu_R$)
at neutrino oscillations (lepton number change on two unites).
Transition Majora neutrino $\nu_L$ into $(\nu_R)^c$ (i.e., $\nu_L
\to (\nu_R)^c$) at oscillations is unobserved since it is supposed
that mass of $(\nu_R)^c$ is very big.

\section{Conclusion}

Usually it is supposed that Majorana neutrino produced in the
superposi\-tion state $\chi_L = \nu_L + (\nu_L)^c$ and then
follows the neutrinoless double beta decay. But since weak
interactions are chiral invariant then the neutrino at production
has definite helicity (i.e.,  $\nu_L$ and $(\nu_L)^c$ neutrinos
are separately produced and then neutrino is not in the
superposition state). This helicity cannot change after production
without any external interactions. Thus we see that for unsuitable
helicity the neutrinoless double $\beta$ decay is not possible
even if neutrino is a Majorana particle. Also transition of
Majorana neutrino into antineutrino at their oscillations is
forbidden since helicity in vacuum holds. Then only possibility to
prove that neutrino is a Dirac but not Majorana particle is
detection transition of $\nu_L$ neutrino into (sterile)
antineutrino $\bar\nu_R$ (i.e., $\nu_L \to \bar\nu_R$) at neutrino
oscillations. Transition Majora neutrino $\nu_L$ into $(\nu_R)^c$
(i.e., $\nu_L \to (\nu_R)^c$) at oscillations is unobserved since
it is supposed that mass
of $(\nu_R)^c$ is very big. \\

\end{document}